\documentclass[runningheads]{svmult}

\usepackage{makeidx}
\usepackage{graphicx}

\usepackage{subeqnar}

\usepackage{multicol}

\usepackage{physprbb}

\begin{document}
\title*{Fundamental units: physics and metrology}

\toctitle{Fundamental units: physics and metrology}

\titlerunning{Fundamental units: physics and metrology}

\author{L.B. Okun
}
\authorrunning{L.B. Okun}

\institute{ITEP, Moscow, 117218, Russia, E-mail: okun@heron.itep.ru}

\label{04}

\maketitle

\begin{abstract}
The problem of fundamental units  is
discussed in the context of achievements of both theoretical
physics and modern metrology. On one hand, due to fascinating
accuracy of atomic clocks, the traditional
macroscopic standards of metrology (second, metre, kilogram) are
giving way to standards based on fundamental units
 of nature: velocity of light $c$ and
quantum of action $h$. On the other hand, the poor precision of
gravitational constant $G$, which is widely believed to define the
``cube of theories'' and the units of the
future ``theory of everything'', does not allow to use $G$ as a
fundamental dimensional constant in metrology. The electromagnetic
units in SI are actually based on concepts of prerelativistic
classical electrodynamics such as ether, electric permitivity and
magnetic permeability of vacuum. Concluding remarks are devoted to
terminological confusion which accompanies the progress in basic
physics and metrology.
\end{abstract}

\section{Introduction}

The problem of fundamental units  has
many facets, three of which seem to be most important:
theoretical, experimental and technological. At present they are
inseparable. Theory, the so called Standard Model,
 formulates basic physical laws and
mathematical methods of their application. Theoretical laws were
established and continue to be established and tested on the basis
of ingenious experiments and astronomical observations of higher
and higher accuracy for an expanding space of parameters.

Precision experiments and observations, in their turn, are
unthinkable without modern high technologies, including lasers and
computers. These technologies are indissolubly  connected with
metrology -- creation, perfection and unification of standards of
physical units, while metrology is widely using the results of
such theories as quantum mechanics, relativity theory,
electrodynamics, condensed matter theory etc. Thus the circle is
closed.

The situation is additionally complicated by the fact that the
Standard Model is not a complete theory. It
has many unsolved problems. Perhaps the most burning is the
problem of existence of fundamental scalar particles (higgses),
responsible for the masses of all fundamental particles (from the
heaviest one -- $t$-quark to the lightest of the three neutrinos).
We still do not understand the role of the three families of
leptons and quarks. It would be too naive to think that their only
justification is CP-violation. We still lack a successful theory
unifying electroweak and strong interactions. The hypothesis of
existence of moderately violated supersymmetry -- symmetry between
fundamental bosons and fermions is still not confirmed by
experiments. Mathematical constructs of the type of superstrings
and M-theory which in the beginning were considered as attempts to
unify quantum gravity with electroweak and strong interaction, as
time goes by, withdraw into a separate field of mathematics, whose
practioneers do not promise any applications to the real physical
world. The situation might become drastically different if
manifestations of extra space dimensions are discovered; in
particular if laws of gravity turned out to change at TeV scale.

Sections 2-6 are devoted to the history of units based on $c$,
$h$, $G$. Sections 7-9 deal with the units based on $c$, $h$, $e$
and precision frequency measurements. Section 10 compares the
Gaussian units and SI units. It is argued that while the latter
are more convenient for practical purposes, the former allow to
better understand the basic notions of modern physics. Therefore
the use of both systems of units should be allowed in physics
textbooks. Section 11 contains concluding remarks.

\section{Fundamental parameters and units}

The essence of theoretical physics is expressed by dimensionless
equations for dimensionless quantities. However one cannot do
experimental physics (and to teach physics) without dimensional
quantities and dimensional units.

In what follows we shall refer to dimensionless fundamental
constants such as $e^2/\hbar c$, or $m_e/m_p$ as fundamental
parameters (here $e$ is the
electron charge, $\hbar$ is the reduced quantum of action ($\hbar
= h/2\pi$) and of angular momentum, $c$ -- velocity of light in
vacuum, $m_e$ and $m_p$ are masses of electron and proton
respectively).

In the absence of established terminology we shall refer to
dimensional fundamental constants as fundamental units. Examples
of units: $c$ (for velocity), $\hbar$ (for action and angular
momentum). According to our definition $G$ is also a fundamental
unit (though indirectly).

\section{Planck units }

When in 1899-1900 Planck discovered $h$ \cite{1}, he used this
discovery to introduce universal units, which at present are
written in the form
\begin{equation} l_P = \hbar/m_P c \; , \;\;
t_P = \hbar/m_P c^2 \; , \;\; m_P = (\hbar c/G)^{1/2} \;\; ,
\label{041}
\end{equation}
where $G$ is Newton's gravitational constant.

Planck derived his units by using dimensional order of magnitude
relations:
\begin{equation}
c = \frac{l_P}{t_P} \; , \;\; \frac{Gm_P^2}{l_P} = m_P c^2 \; ,
\;\; \frac{Gm_P^2}{l_P}t_P = \hbar \;\; .  \label{042}
\end{equation}

He was inspired by the idea that his units are universal (contrary
to ``handcrafted'' earthbound  ordinary units -- meter, second,
gram): they are the same at any far away corner of the universe.

Planck also considered as universal the Planck temperature $T_P =
m_P c^2/k$. But Boltzman's $k$ is not a universal unit, it is a
conversion factor: $ k = 8.6 \cdot 10^{-5}$ eV/K (hint: $\hbar
\omega/kT$).

\section{$\mbox{\boldmath$c, h, G$}$ -- units}

From the point of view of the future ``theory of everything'' it
is natural to use $c, h, G$ as fundamental dimensional constants.

In 1928 Gamov, Ivanenko and Landau \cite{2} considered the theory
``of the world as whole'' in terms of dimensional fundamental
constant $c$, $h$, $G$. In 1928 Landau was 20 years old, Gamow and
Ivanenko -- 24 (see Figs.~ 1,2). They had written the paper
``World Constants and Limiting Transitions'' (see Fig.~3) as a
humorous birthday present to their friend, a young lady. None of
them ever referred this paper in their subsequent publications.
But the ideas of the paper were fundamental. In 1936 Bronstein
\cite{3} worked at a theory in which all three constants are
finite. It was one of the first papers on relativistic quantum
gravity. In 1967 ideas of refs. \cite{2,3} were presented in the
form of a cube (see Figs.~4,5) by Zelmanov \cite{4}. Later on it
was further developed by others \cite{5,6}.

The vertices of this cube represent nonrelativistic mechanics
(NM), nonrelativistic gravity (NG), nonrelativistic quantum
mechanics (QM), special relativity (SR), quantum field theory
(QFT), general relativity (GR) and finally relativistic quantum
gravity (QGR) or theory of everything (TOE).

The cube, made of units, is ``endowed'' with dimensionless
parameters like $\alpha, \alpha_s$, mixing angles, mass ratios,
etc. Their values are expected to follow from TOE. Similar to the
cube is ``dimensional pyramid''
(Kuchar \cite{7}, Sanchez \cite{8}) with 4 vertices and 4 planes,
Fig. 4.

Note that Einstein tried to build a unified theory of electricity
and gravity (``TOE'') in the left-hand vertical plane of the cube,
without Quantum Mechanics, without $\hbar$.

Planck units allow one to deal in the equations of TOE only with
dimensionless functions of dimensionless variables and
dimensionless parameters of the type $\alpha = e^2/\hbar c$ or
$m_e/m_p$. Conceptually Planck units are excellent, but
practically they have serious shortcomings, caused by $G$, the
same $G$ which allows to bring gravitation and cosmology into the
realm of quantum phenomena. Thus the source of strength at the
same time turns out to be a source of weakness.

\section{Planck units are impractical}

The obvious shortcoming of Planck units is that they differ by
many orders of magnitude from atomic units commonly used in
physics. Their values are natural for the early universe and TOE,
but not for mundane physics: $$ l_P = 10^{-35} \; {\rm m} \; ,
\;\; t_P = 10^{-43} \; {\rm s} \; , \;\; E_P = m_P c^2 = 10^{19}
\; {\rm GeV} \;\; . $$

The energy which corresponds to the Planck mass is unattainable by
accelerators even of the remotest future. (Note, however, that it
is only a few orders of magnitude larger than the grand
unification scale of electroweak and strong interactions.) The
Planck units of length and time are vanishingly small compared
with atomic units. Of course the huge powers of ten are not
frightful by themselves. As is well known, atomic units also
differ by many orders of magnitude from SI units, which does not
prevent atomic standards to be the base of modern metrology.

Much more essential is another shortcoming of Planck units, which
stems from the fact that $G$ is known with rather poor accuracy
(of order of $10^{-3}$, by five -- four orders worse than those of
$c$ and $h$, and by 12 orders worse than the precision of atomic
clocks). Thus it is impossible to use the Planck units as
standards in modern precision physics and technology.

\section{Units of Stoney}

The use by Planck of $G$ as a basis for defining the unit of mass
was caused by absence at the beginning of the 20th century of
another natural, not ``handcrafted'', candidate for the unit of
mass. In that respect Planck's universal units resemble the
universal units suggested 30 years earlier by Irish physicist
Stoney (1826 - 1910), secretary of Irish Royal Society (see
Figs.~7,8). By studying electrolysis, he was the first who
measured the value of elementary charge $e$ and introduced into
physics the term ``electron'' for the carrier of this charge (in
modern terminology it is ion). From $e,c,G$ ~ Stoney \cite{7}
constructed in 1870 - 1880 universal units with dimensions of
length, time and mass:
\begin{equation}
l_{\rm S} = e\sqrt G/c^2 \; , \;\; t_{\rm S} = e\sqrt G/c^3 \; ,
\;\; m_{\rm S} = e/\sqrt G \;\; , \label{043}
\end{equation}
which he derived from dimensional equations: $$c = l_{\rm
S}/t_{\rm S} \; , \;\; e^2 = G m_{\rm S}^2 \; , \;\; e^2/l_{\rm S}
= m_{\rm S} c^2 \;\; . $$

Let us note that units of Stoney are only by a factor
$\sqrt\alpha$ smaller than those of Planck.

Stoney's units look ``tailored'' for
Einstein's unified theory. Constants $e$, $c$, $G$ contain the
gist of classical electrodynamics and gravity. There is no $\hbar$
in them. Comparison with $c$, $\hbar$, $G$ shows that $\hbar$ is
brought into Stoney's set of constants ``through the back door of
$\alpha$''. Therefore $e$, $c$, $G$ do not form a cube of theories
 with its limiting transitions considered
by Gamov, Ivanenko and Landau \cite{2}.

\section{Atomic clocks and $\mbox{\boldmath$c$}$}

During the 20th century the situation with standard of mass (time,
length) has changed drastically. The fundamental identity of
elementary particles and hence of atoms produced many candidates
for standard of mass, known with much, much better precision than
$G$. Thus, from the point of view of dimensions the necessity to
use $G$ disappeared. However from the point of view of unifying
physics the Planck units became even more attractive.

Let us now look at two other fundamental constants: $c$ and $h$.

Let us start from $c$ and the frequencies of light and radio
waves. In the second half of the 20th century  physicists learned
how to measure them in a digital way by counting the number of
crests. This raised the accuracy of atomic (Cesium-133) clocks
(first suggested by I.~Rabi in 1945) to the level of 1 second in
300 years (NBS, 1955). (Now this has become 1 second in $20 \cdot
10^6$ years: LPTF, NIST, PTB.) But even the first figure was
sufficient for the introduction into SI of an atomic unit of a
second (in 1967):

``1 s $= 9\;192\;631\;770$ periods of radiation in transition
between levels of hyperfine splitting of the atomic ground state
of Cs-133''.

This, together with the independence of the velocity of light on
its frequency, impelled Bay et al. \cite{8} to suggest, instead of
unit of length (meter), to use as the basic unit the unit of
velocity, namely the velocity of light $c$. In 1983 the definition
\begin{equation}
c = 299\;792\;458 \; {\rm m/s} \label{044}
\end{equation}
was introduced in SI. The traditional standard of length gave way
to the new standard based on the value of the velocity $c$. This
velocity is defined as a number without uncertainty. Further
improvements of experiments which measured $c$ would mean further
improvement of the realization of the meter. An international
report ``Practical realization of the definition of the metre,
including recommended radiations of other optical frequency
standards'' (2001) was published by T.~Quinn in 2003 \cite{9}.
(Note that both spellings ``metre'' and ``meter'' are used in the
literature, the former in metrology, while the latter one in
physics.)

Further progress in accuracy of atomic clocks
is connected with passing from microwave to
optical frequencies \cite{10,11}.

\section{Towards kilogram based on $\mbox{\boldmath$h$}$}

Thus metrology made two momentous steps in the direction of
fundamental physics: the place of macroscopic clocks and ruler
(the famous rod at BIMP, in Sevre, near Paris) became occupied by
the velocity of light and by atoms of Cs-133. There remains now
only one macroscopic standard -- the kilogram at Sevre. The
prospect of expressing it through the quantum of action $h$ is
connected with precision measurements in atomic and condensed
matter physics. There are many promising quantities which are good
candidates for such measurements. I shall touch upon only one
project which is connected with two outstanding discoveries in
condensed matter physics: the Josephson effect
\cite{12} (Nobel Prize 1973) and the von
Klitzing effect \cite{13} (Nobel Prize
1985).

Josephson theoretically predicted the existence of a supercurrent
and its remarkable properties. A supercurrent is a current of
Cooper pairs tunneling through an insulator separating two
superconductors. A supercurrent can exist without external
voltage. An external voltage $V$ creates an alternating
supercurrent of frequency $\nu$. The steps in $V$ are given by the
relation:
\begin{equation}
V(n) = \nu n/K_J \;\; , \label{045}
\end{equation}
where $n$ is an integer, while the coefficient $K_J$ is universal
and is called the Josephson constant. It is reproduced in various
experiments with unprecedented accuracy and is determined only by
the ratio of fundamental constants:
\begin{equation}
K_J = 2e/h \;\; . \label{046}
\end{equation}

The effect, discovered by von Klitzing, is called the quantum Hall
effect. This effect shows that there exists in Nature a universal
electric resistance, one which can be expressed in terms of
fundamental constants.

As is well known, the ordinary Hall effect occurs in a solid
conductor (or semiconductor) with density of current {\bf j} in a
magnetic field {\bf H} which produce an electric field {\bf E}
(with voltage $V_H$) orthogonal
both to {\bf j} and {\bf H}.

The quantum Hall effect was discovered in a two-dimensional
electron system separating two parts of a silicon field transistor
at very low temperature ($< 4$ K) and very strong magnetic field
($\sim 14$ Tesla). It was established that the Hall resistance
\begin{equation}
R_H = V_H/I \;\; , \label{047}
\end{equation}
where $I$ is the total current, has quantum jumps:
\begin{equation}
R_H(n) = \frac{R_K}{n} \;\; , \label{048}
\end{equation}
where $n$ is an integer, while $R_K$ is the von Klitzing constant:
\begin{equation}
R_K = h/e^2 \;\; . \label{049}
\end{equation}

It is obvious that
\begin{equation}
h = 4K_J^{-2} R_K^{-1} \;\; . \label{0410}
\end{equation}
This permits measurement of $h$ using macroscopic apparatus. A
special two-story-high watt balance compared
electrical and mechanical forces:
\begin{equation}
VI/v = mg \;\; , \label{0411}
\end{equation}
where $m$ is the measured mass of a body, $g$ -- local
gravitational acceleration, $V$ -- the voltage in a coil moving
with a vertical velocity $v$ in a magnetic field, while $I$ is the
current in the same coil, this time fixed in the same magnetic
field. By calibrating $V$ and $V/I$ through the Josephson and von
Klitzing effects Williams et al. \cite{14} succeeded in connecting
$h$ and the kilogram within uncertainty $8.7 \cdot 10^{-8}$.

It is hoped that in the not too distant future this accuracy might
be improved by an order of magnitude, which would allow to use the
watt balance for gauging the standards of mass and thus get rid of
the Sevres kilogram and to define the value of $h$. As a result
the value of $h$ would have no uncertainties in the same way as it
occurred with $c$. Thus fundamental units of nature $c$ and $h$
would become fundamental SI units of metrology.

\section{Kilogram as frequency $\mbox{\boldmath$\nu_K$}$}

Another definition of the kilogram has been suggested \cite{15} on
the basis of equations
\begin{equation}
E = h\nu \;\; , \label{0412}
\end{equation}
\begin{equation}
E = mc^2 \;\; \mbox{\rm :} \label{0413}
\end{equation}
``The kilogram is the mass of a body at rest whose equivalent
energy equals the energy of collection of photons whose
frequencies sum to $13.5639274 \times 10^{49}$ hertz''.

This definition should be taken with a grain of salt. The combined
use of equations (\ref{0412}) and (\ref{0413}) implies that a
photon of frequency $\nu$ has mass $h\nu/c^2$. This implication
persists in spite of the words ``equivalent energy''. The words
``the mass of the body at rest'' imply that mass is not Lorentz
invariant, but depends on the velocity of a reference frame. It
would be proper to replace equation (\ref{0413}) by
\begin{equation}
E_0 = mc^2 \;\; , \label{0414}
\end{equation}
where $E_0$ is the rest energy (see e.g. ref. \cite{16}). But then
it would take some additional considerations in order to define
the frequency $\nu_K$ corresponding to one kilogram. In particular
massive atoms emitting and absorbing photons should be taken into
account. From practical point of view the measurement of
``frequencies sum'' of order $10^{50}$ hertz is by eight orders of
magnitude more difficult than that of the Planck frequency $\nu_P
= 1/t_P$.

\section{Electromagnetism and Relativity}

Electromagnetism -- the kinship of electricity and magnetism,
discovered in 1820 by Oersted, rather soon became the foundation
of Amp\'{e}re's  electrodynamics. The development of the latter by
Faraday and other outstanding physicists culminated in 1873 in the
Treatise of Maxwell \cite{1''} who linked electric currents with
electric and magnetic fields and with the properties of light.
None of these great physicists knew the genuine nature of the
phenomena. Maxwell considered vacuum filled with ether; the
carriers of charges were unknown to him. The electromagnetic field
was described by four vector quantities: electric field {\bf E},
electric induction (or
displacement) {\bf D}, magnetic field {\bf H},
and magnetic induction (or flux density) {\bf B}.

On the basis of these notions practical units (such as volt,
ampere, coulomb, joule) were introduced by International
Electrical Congresses in 1880s. The electric permitivity
$\varepsilon_0$
and
magnetic permeability $\mu_0$
ascribed by
Maxwell to the ether were accepted by the community of engineers
and physicists: ${\rm \bf D} = \varepsilon_0 {\rm\bf E}$, ${\rm\bf
B} = \mu_0 {\rm\bf H}$. In the middle of the 20th century these
practical units became the basis of the Syst\`{e}m International
d'Unit\'{e}s (SI).

The end of the 19th and beginning of the 20th century were marked
by great successes in understanding and applying classical
electrodynamics. On practical side it was the use of electric
currents in industry, transport and radio communications. On
theoretical side it was unification of electrodynamics, optics and
mechanics in the framework of special relativity \cite{13''}.

According to special relativity, the 4-radius vector is $x^i =
(ct, {\rm\bf r})$ ($i = 0,1,2,3$), the 4-momentum vector is $p^i =
(E/c, {\rm\bf p})$, the 4-potential of electromagnetic field $A^i
= (\varphi, {\rm\bf A})$, the density of the 4-current $j^i = (c
\rho, {\rm\bf j})$, where ${\rm\bf j} = \rho {\rm\bf v}$, and
$\rho = e\delta({\rm\bf r} - {\rm\bf r}_a)$, $e$ is electric
charge. (The current $j^i$ is consistent with the definitions of
$p^i$ and $A^i$, due to an appropriate coefficient $c$ in front of
$\rho$. The source of the field, the charge, is pointlike.
Otherwise there appears a problem of the field inside the
finite-size cloud of charge.) The upper index $i$ of a 4-vector
indicates a contravariant 4-vector; a lower index $i$ indicates
covariant 4-vector, its space components have minus sign. Raising
or lowering of indices is done with the diagonal metric tensors
$g^{ik}$ or $g_{ik}$ respectively.

The 3-vectors {\bf E} and {\bf H} are components of the 4-tensor
of electromagnetic field
\begin{equation} F_{ik} = \frac{\partial
A_k}{\partial x^i} - \frac{\partial A_i}{\partial x^k} \;\; .
\label{0415}
\end{equation}

The tensors $F_{ik}$ and $F^{ik}$ can be represented by matrices:
\begin{equation}
F_{ik} = \left( \begin{array}{cccc} 0& E_1 & E_2 & E_3 \\ -E & 0 &
-H_3 & H_2 \\ -E_2 & H_3 & 0 & -H_1 \\ -E_3 & -H_2 & H_1 & 0
\end{array} \right) \;\; , \label{0416}
\end{equation}
and
\begin{equation}
F^{ik} = \left( \begin{array}{cccc} 0& -E_1 & -E_2 & -E_3 \\ E & 0
& -H_3 & H_2 \\ E_2 & H_3 & 0 & -H_1 \\ E_3 & -H_2 & H_1 & 0
\end{array} \right) \;\; , \label{0417}
\end{equation}
respectively, or in a condensed form:
\begin{equation}
F_{ik} = ({\rm\bf E}, {\rm\bf H}) \;\; , \label{0418}
\end{equation}
\begin{equation}
F^{ik} = (-{\rm\bf E}, {\rm\bf H}) \;\; . \label{0419}
\end{equation}

This 4-tensor is obviously antisymmetric. From the definition of
$F_{ik}$ it follows that dimensions of {\bf E} and {\bf H} are the
same: $[{\rm\bf E}] = [{\rm\bf H}]$.

The field equations have the form in Gaussian units:
\begin{equation}
 \frac{\tilde F^{ik}}{\partial
x^k} = 0 \;\; , \label{0420}
\end{equation}
\begin{equation}
\frac{\partial F^{ik}}{\partial x^k} = -\frac{4\pi}{c} j^i \;\; .
\label{0421}
\end{equation}
 Here
\begin{equation}
  \tilde F^{ik} = \varepsilon^{iklm} F_{lm} \;\; , \label{0422}
  \end{equation}
where $\varepsilon^{iklm}$ is fully antisymmetric tensor
($\varepsilon^{0123} = +1$).

The equation describing the motion of charge in the
electromagnetic field is given by
\begin{equation}
 \frac{d{\rm\bf p}}{dt} = e
{\rm\bf E} + \frac{e}{c}[{\rm\bf v H}] \;\; , \label{0423}
\end{equation}
where
\begin{equation}
{\rm\bf v} = \frac{{\rm\bf p}c^2}{E} \;\; . \label{0424}
\end{equation}

Note that according to special relativity there is no ether,
$\varepsilon_0 \equiv \mu_0 \equiv 1$, and the strength of
magnetic field in vacuum ${\rm\bf H}$ has the same dimension as
that of ${\rm\bf E}$, the identity of $\varepsilon_0 \equiv \mu_0
\equiv 1$ immediately follows from the fact that the same $e$
determines the action of the charge on the field and of the field
on the charge. (See expression for the action in ref. \cite{6''},
eq. (27.6).) Thus, there is no need to consider ${\rm\bf B}$ and
${\rm\bf D}$ in the case of vacuum. In classical electrodynamics
they appear only in the continuous media due to polarization of
the latter \cite{7''}.

In a number of classical monographs and textbooks on classical
electrodynamics ${\rm\bf E}$ and ${\rm\bf H}$ are consistently
used for description of electric and magnetic fields in vacuum
with $\varepsilon_0 \equiv \mu_0 \equiv 1$ (\cite{13''},
\cite{6''}, \cite{10''}, \cite{11''}, \cite{12''}). Their authors
use Gaussian or Heaviside-Lorentz (with $1/4\pi$ in Coulomb law)
units.

Many other authors use {\bf B} instead of {\bf H}, sometimes
calling {\bf B} magnetic field and sometimes -- magnetic induction
in vacuum \cite{8''}. Most of them use the SI units, according to
which $\varepsilon_0$ and $\mu_0$ are dimensional: $\mu_0 = 4\pi
\cdot 10^{-7}$HA$^{-2}$, $\varepsilon_0 \mu_0 = c^2$, where H is
henry, while A -- ampere. The classical electromagnetic fields in
vacuum are described by four physical quantities ${\rm\bf D}$,
${\rm\bf H}$ and ${\rm\bf E}$, ${\rm\bf B}$, all four of them
having different dimensions at variance with the spirit of special
relativity.\footnote{Sometimes one can hear that the identity
$\varepsilon_0 \equiv \mu_0 \equiv 1$ is similar to putting $c=1$,
when using $c$ as a unit of velocity. However this similarity is
superficial. In the framework of special relativity one can use
any unit for velocity (for instance, m/s). But the dimensions and
values of $\varepsilon_0$ and $\mu_0$  are fixed in SI.} In that
respect vacuum is similar to a material body. The SI units are
very convenient for engineers, but not for theorists in particle
physics.

In fact, theorists are not less responsible than metrologists for
the gap between deductive basis of modern physics and mainly
prerelativistic inductive basis of modern metrology. A good
example is the 1935 article \cite{3''} by A.~Sommerfeld and his
book ``Electrodynamics'' based on lectures given in 1933-34
\cite{4''}.

His argument against absolute system (that is based on units of
time, length and mass) was the presence in it of fractional
exponents (for instance from Coulomb law the dimension of charge
is ${\rm g}^{1/2}$ cm$^{3/2} {\rm s}^{-1}$). This argument was not
very compelling in the 1930s and is even less so today. His
argument against Gaussian or Heaviside-Lorentz system was based on
inductive, prerelativistic view on electromagnetism.  Though he
was not quite happy\footnote{``What is especially painful for me
is that the fine structure constant
is no more $e^2/\hbar c$, but
$e^2/4\pi \varepsilon_0 \hbar c$''. Z. Phys. {\bf 36} (1935) 818.}
with the new clumsy expression for fine structure constant
$\alpha$ introduced by him  before the World War I, he kept
insisting on MKSA units and against Gaussian units. His authority
was not the least in the decision to legally enforce after World
War II the SI as the obligatory system of units for all textbooks
in physics.

Coming back to classical electrodynamics let us note that it is
not a perfect theory: it has serious problems at short distances.
To a large part these problems are solved by quantum
electrodynamics (QED). Therefore the latter should be used as a
foundation of a system of electromagnetic units. By the way, QED
is used to extract the most accurate value of $\alpha$ from the
precision measurements of the magnetic moment of electron.

In the framework of QED $\alpha$ is not a constant but a function
of momentum transfer due to polarization of vacuum. Let us stress
that this polarization has nothing to do with purely classical
non-unit values of $\varepsilon_0$ and $\mu_0$.

\section{Concluding remarks}

Mutually fruitful ``crossing'' of fundamental physics and
metrology gives numerous practical applications. One of them
should be specially mentioned: the use of general relativity in
global positoning system \cite{17',17}.

Remarkable achievements of metrology are not always accompanied by
elaboration of adequate terminology. Here we will mention only a
few of widely spread delusions.

The choice of $c$ as a unit of velocity leads many authors to the
false conclusion that $c$ should be excluded from the set of
fundamental units. They insist that $c=1$, because $c$ in units of
$c$ is equal 1. (The same refers to $h$ in units of $h$.) But
number 1 is not a unit of measurement,
because such units are always dimensional. Equations $c, h =1$ are
simply wide spread jargon. Some authors go even further by
identifying space and time. (A detailed discussion can be found in
ref. \cite{18}.)

The number of physical units is not limited. When solving a given
problem the choice of units is determined by considerations of
convenience. However from the point of view of ``the world as a
whole'' $c$, $h$ and $G$ (or instead of $G$ some other quantity
representing gravity) are definitely singled out as fundamental
dimensional constants. Of course they must be accompanied by a
number of dimensionless parameters. But the number of fundamental
units could not be less than three \cite{18}.

The inclusion of candela into the set of base units (see Fig.~9)
seems to be unconvincing from the point of view of physics. Of
course, practically it is convenient to use it when discussing the
brightness of light. But it does not look logical to put it on the
same footing as units of length, time and mass.

As SI is imposed on the physics literature by governmental laws,
the obligatory usage in textbooks of such notions as permitivity
$\varepsilon_0$ and permeability $\mu_0$ of vacuum, makes it
difficult to appreciate the beauty of the modern electrodynamics
and field theory. It corresponds to the prerelativistic stage of
physics.

This list can be extended, but it seems that the above remarks are
sufficient for a serious discussion. The metrological institutes
and SI are of great importance for science and technology.
Therefore the metrological legal documents should be to a greater
degree based on modern physical concepts. Especially they should
give more freedom to the usage of Gaussian and Heaviside-Lorentz
systems of units in the textbooks.

\section{Acknowledgements}

I would like to thank A.~Clairon, H.~Fritzsch, J.~Jackson,
S.~Karshenboim, N.~Koshelyaevsky, H.~Leutwyler, N.~Sanchez,
M.~Tatarenko, V.~Telegdi, Th.~Udem, L.~Vitushkin and H.~Wagner for
fruitful discussions. The work has been partly supported by the
Russian Federal Special Scientific-Technological Program of
Nuclear Physics Fund 40.052.1.1.1112 and by A. von Humboldt award.

\newpage

\section*{Figure captions}

\begin{tabular}{ll}
Fig. 1. &  Meeting in Kharkov, 1928, attended by Gamow, Ivanenko,
and Landau. \\ Fig. 2. & Who is who in Fig.1. \\ Fig. 3. & English
version of the 2002 reprint of the 1928 article by Gamow,
Ivanenko, and Landau. \\ Fig. 4. & Orthogonal axes. At the origin
there is no gravity, no maximal velocity, no quantum \\ & effects.
\\ Fig. 5. & Cube of theories. \\ Fig. 6. & Dimensional pyramid.
\\ Fig. 7. & Cover of the book dedicated to G.J.~Stoney. \\ Fig.
8. & The title page of ``Philosophical Magazine and the beginning
of article \cite{7} by G.J.~Stoney. \\ Fig. 9. & The base units of
the SI, with their present uncertainties of realization, and some
of their \\ & links to atomic and fundamental constants with their
present uncertainties in terms of the \\ & SI. The absence of a
useful quantitative estimate of the long-term stability of the \\
& kilogram, indicated by ?, is reflected in three of the other
base units. The dashed lines to \\ & the kilogram  indicate
possible routes to a new definition. (From the article by
T.J.~Quinn \\ & ``Base Units of the Syst\`{e}me International
d'Unit\'{e}s, their Accuracy, Dissemination and  \\ &
International Traceability'', Metrologia {\bf 31} (1994/95)
515-527.).
\end{tabular}

\newpage
\begin{figure}
\begin{center}
\includegraphics[width=1.0\textwidth]{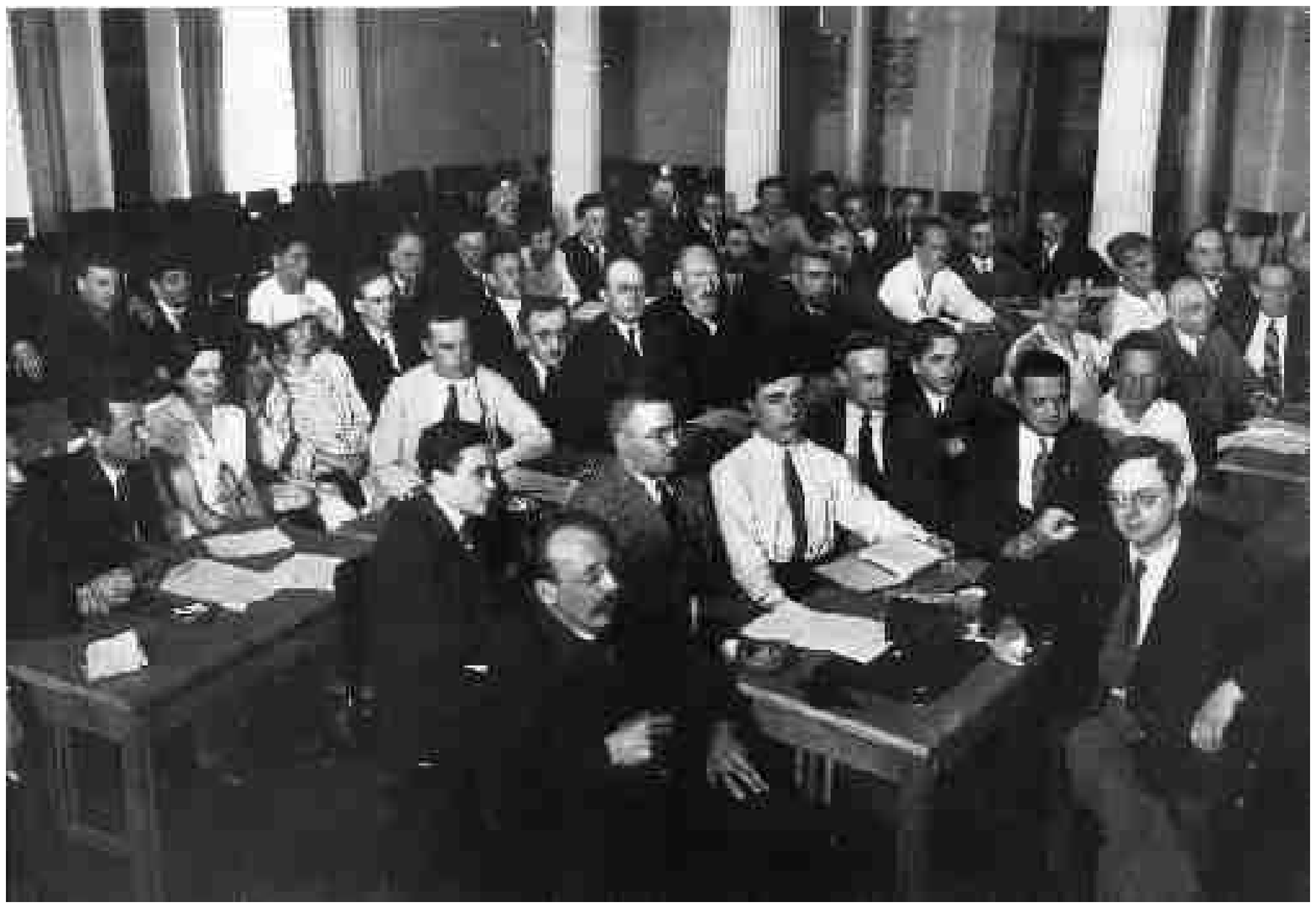}
\caption{Meeting in Kharkov, 1928, attended by Gamow, Ivanenko,
and Landau \label{fig1}} \vspace{2cm}
\includegraphics[width=1.0\textwidth]{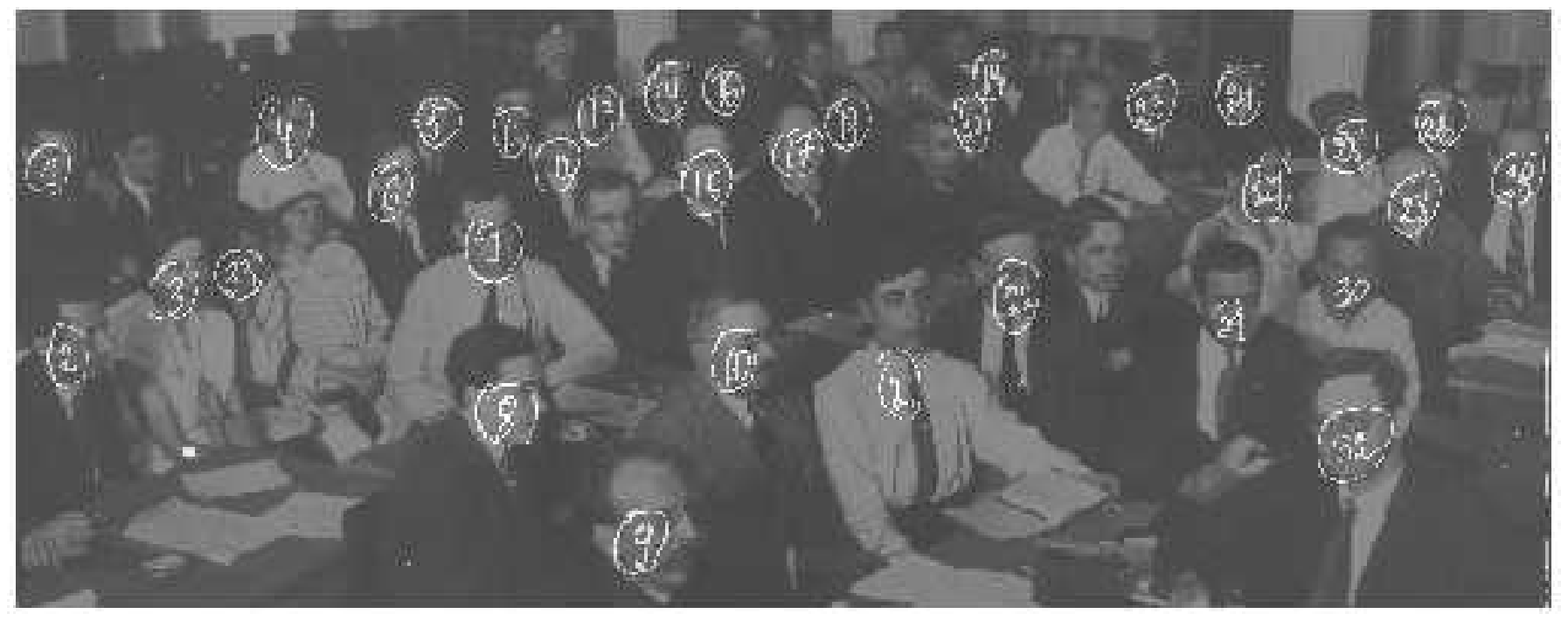}
\caption{Who is who in Fig.1. \label{fig2}} \vspace{1cm}
\begin{tabular}{lll}
1. Dunaev & 12. Kopp  & 23. Vereschagin  \\ 2. Heitler & 13.
Kotsarova  & 24. Slutsky  \\ 3. Arsenyeva & 14. Khalfin & 25.
Gamov \\ 4. Davydov & 15. Efimovich  & 26. Shubnikov
\\ 5. Todorovich(?) & 16. Ogievetsky  & 27. Landau \\ 6. Frish &
17. Grommer  & 28. Shtrumm \\ 7. Bursian & 18. Muskelischvili& 29.
Frenkel
\\ 8. Ivanenko & 19. Korsunsky  & 30. Rosenkevich \\
9. Obreimov & 20. Gorwitz   & 31. Finkelshtein \\ 10. Fock &
 21. Ambartsumian  & 32. Jordan \\ 11. Leipunsky & 22. Mandel & 33. Timoreva
\end{tabular}
\end{center}
\end{figure}

\begin{figure}
\begin{center}
\includegraphics[width=1.0\textwidth]{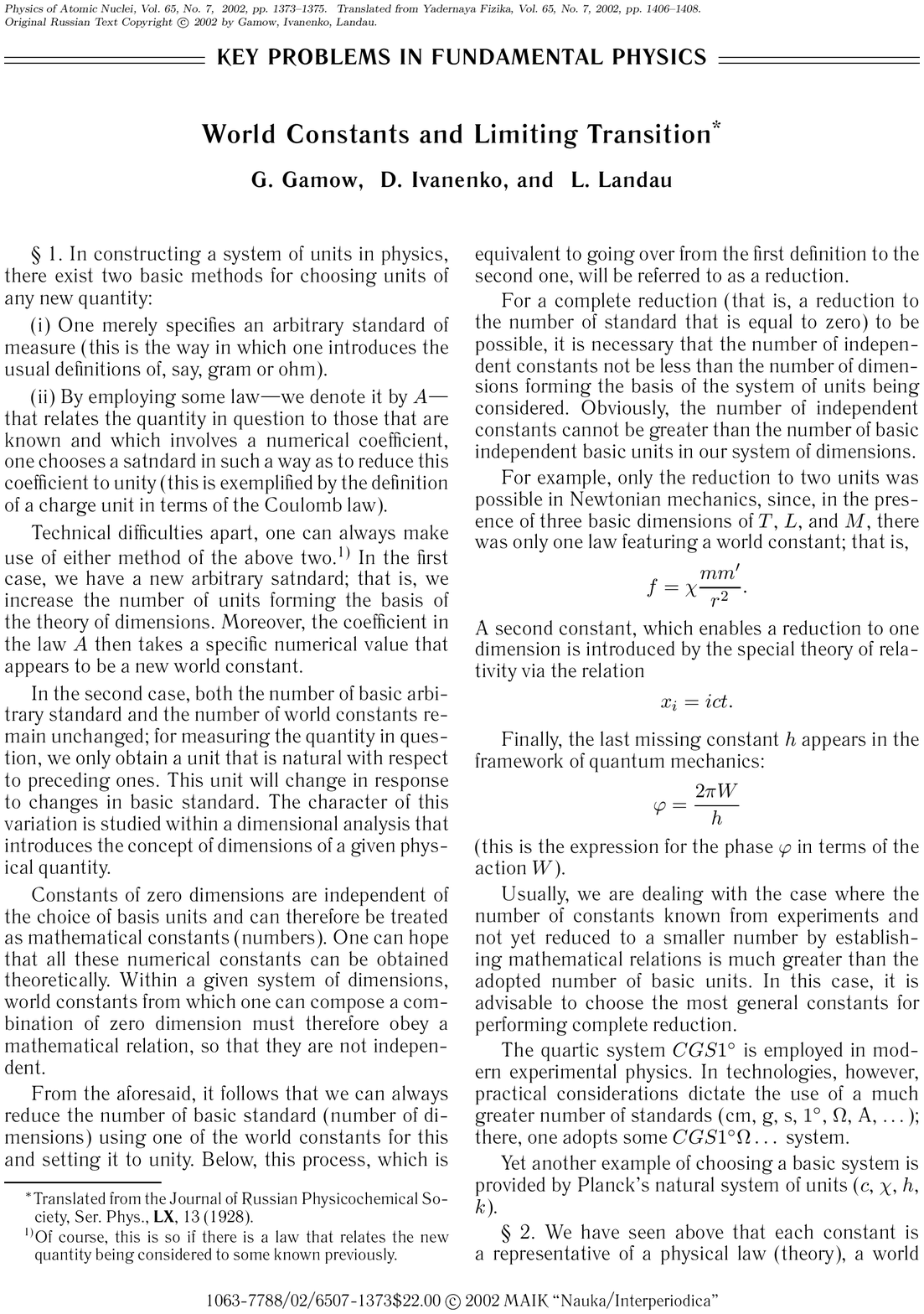}
\caption{English version of the 2002 reprint of the 1928 article
by Gamow, Ivanenko, and Landau. \label{fig3}}
\end{center}
\end{figure}

\begin{figure}
\begin{center}
\includegraphics[width=0.5\textwidth]{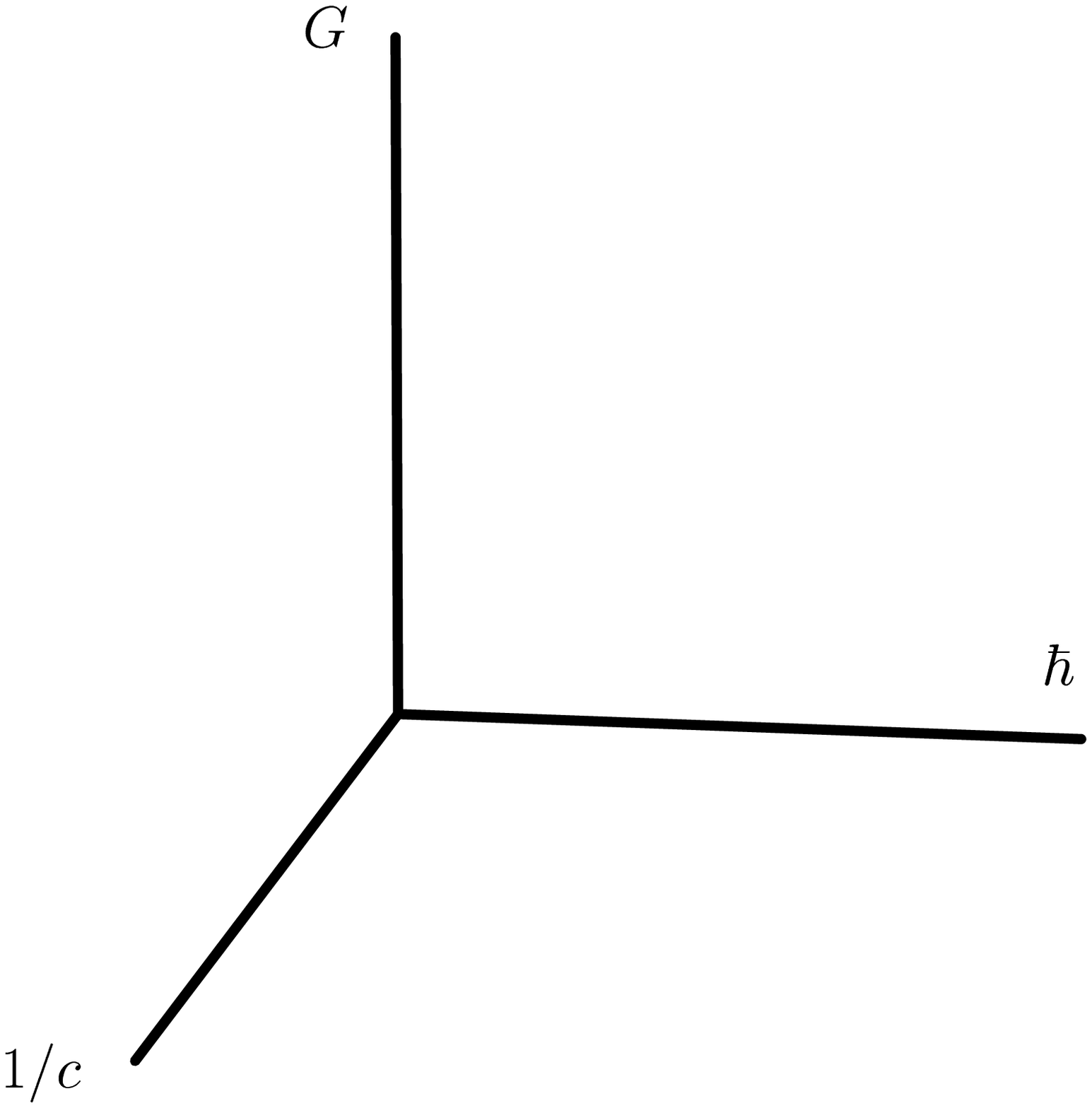}
\caption{Orthogonal axes. At the origin there is no gravity, no
maximal velocity, no quantum effects \label{fig4}}
\end{center}
\end{figure}

\begin{figure}
\begin{center}
\includegraphics[width=0.5\textwidth]{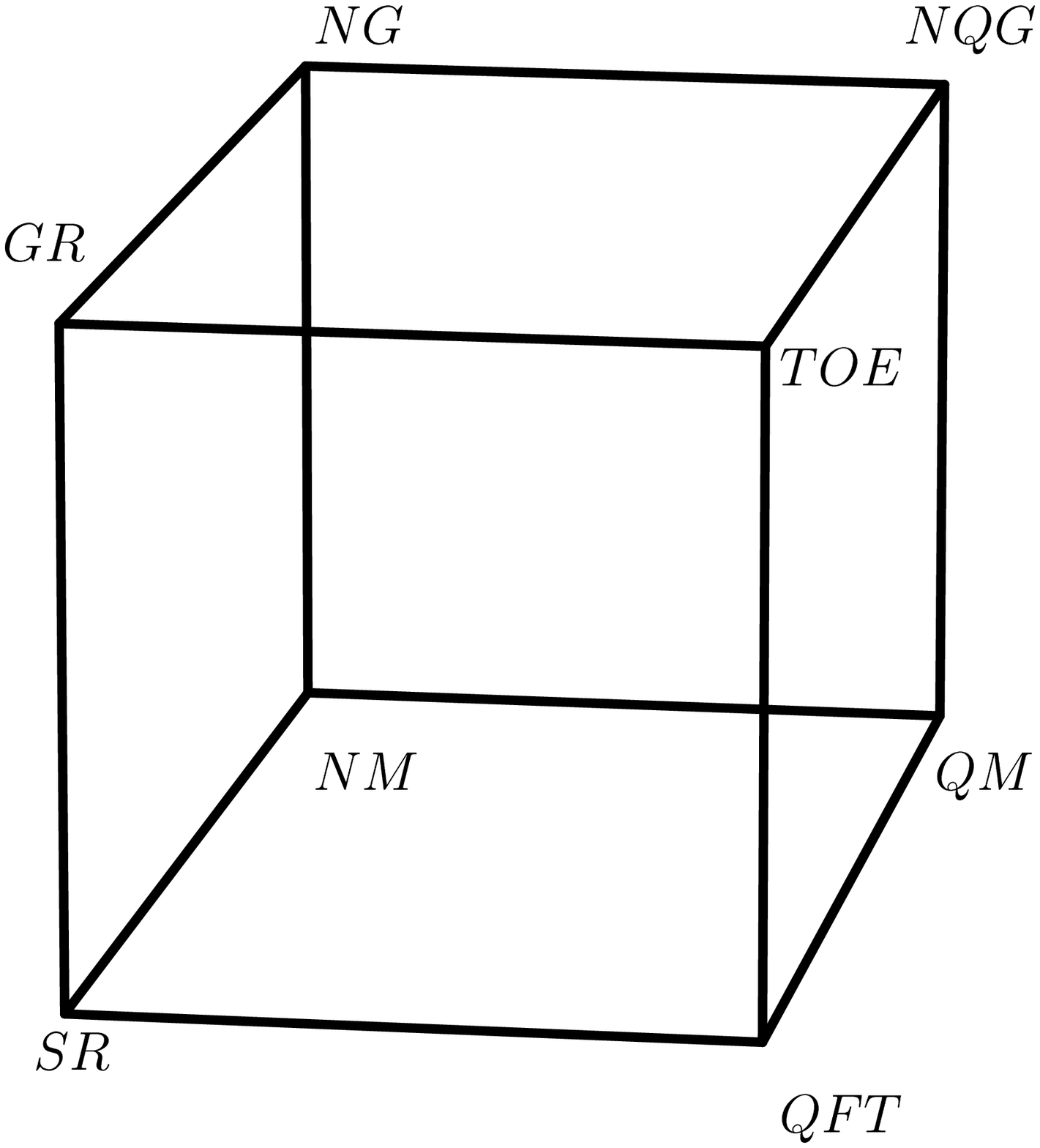}
\caption{Cube of theories \label{fig5}}
\end{center}
\end{figure}

\begin{figure}
\begin{center}
\includegraphics[width=1.0\textwidth]{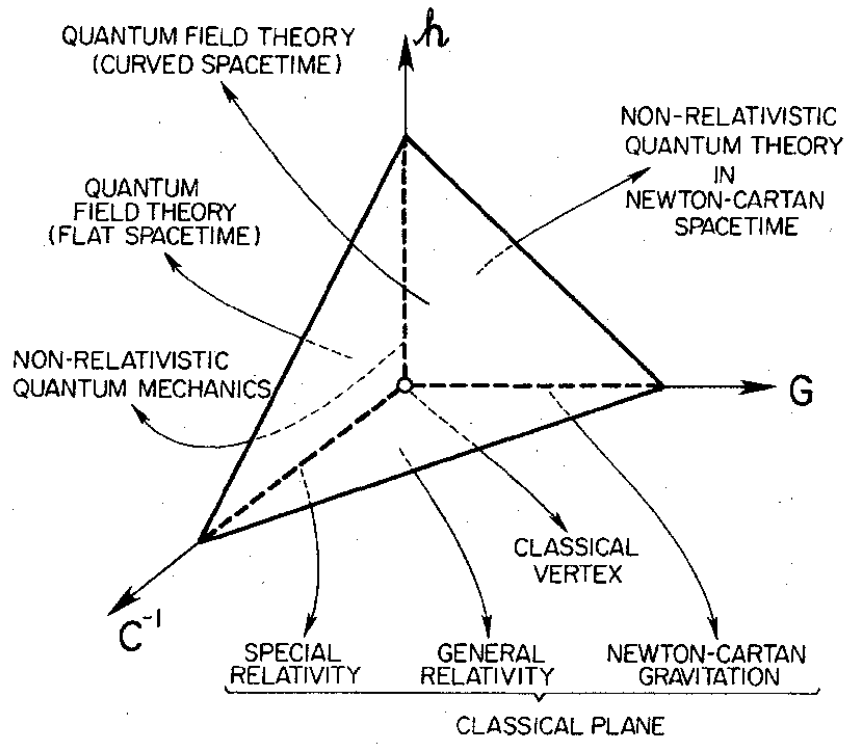}
\caption{Dimensional pyramid  \label{fig6}}
\end{center}
\end{figure}

\begin{figure}
\begin{center}
\includegraphics[width=1.0\textwidth]{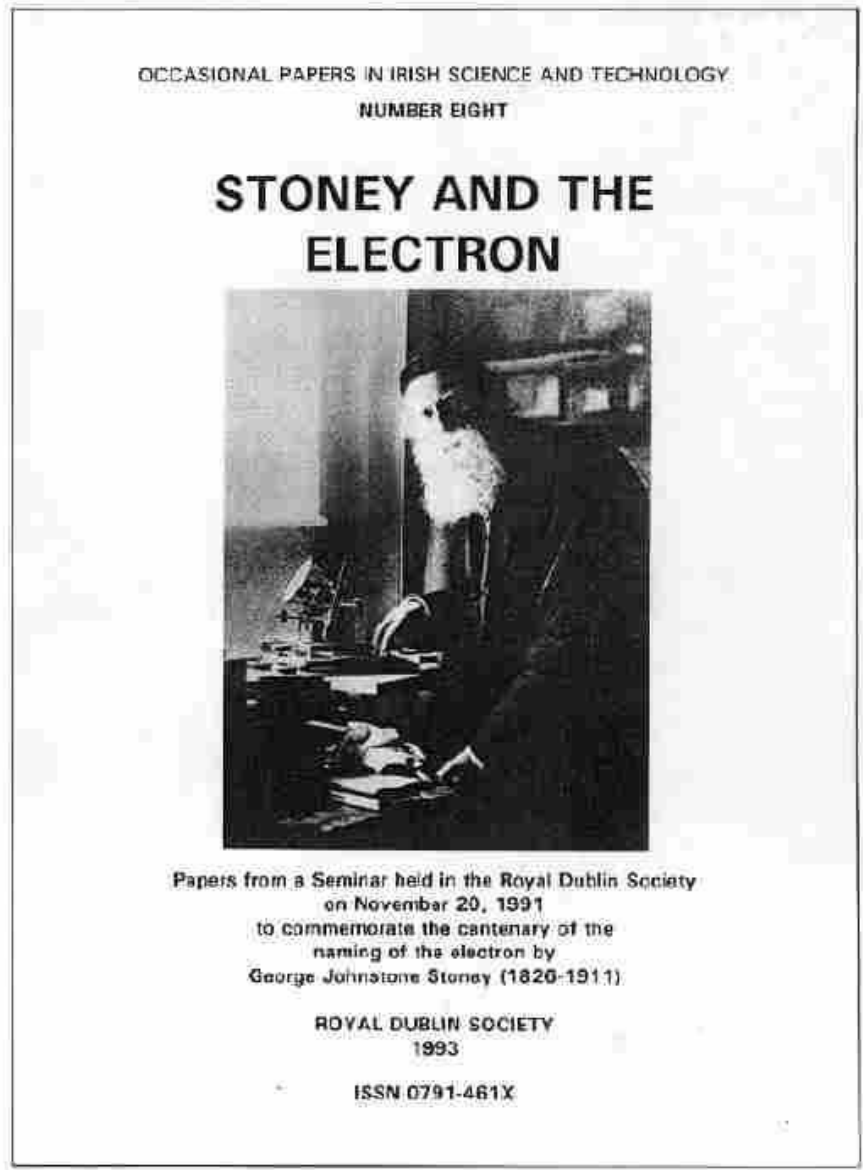}
\caption{Cover of the book dedicated to G.J.~Stoney \label{fig7}}
\end{center}
\end{figure}

\begin{figure}
\begin{center}
\includegraphics[width=1.0\textwidth]{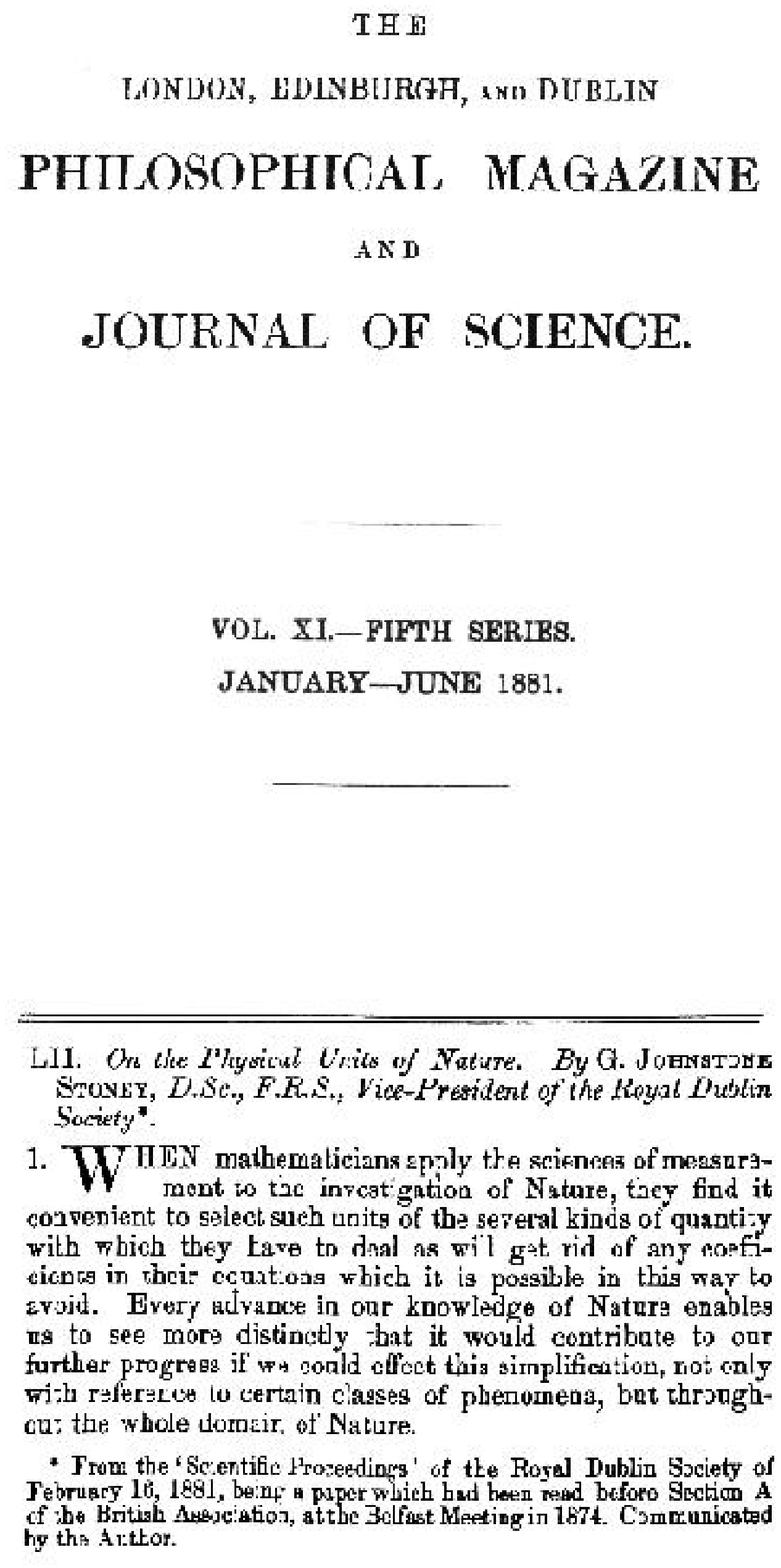}
\caption{The title page of ``Philosophical Magazine and the
beginning of article \cite{7} by G.J.~Stoney \label{fig8}}
\end{center}
\end{figure}

\begin{figure}
\begin{center}
\includegraphics[width=1.0\textwidth]{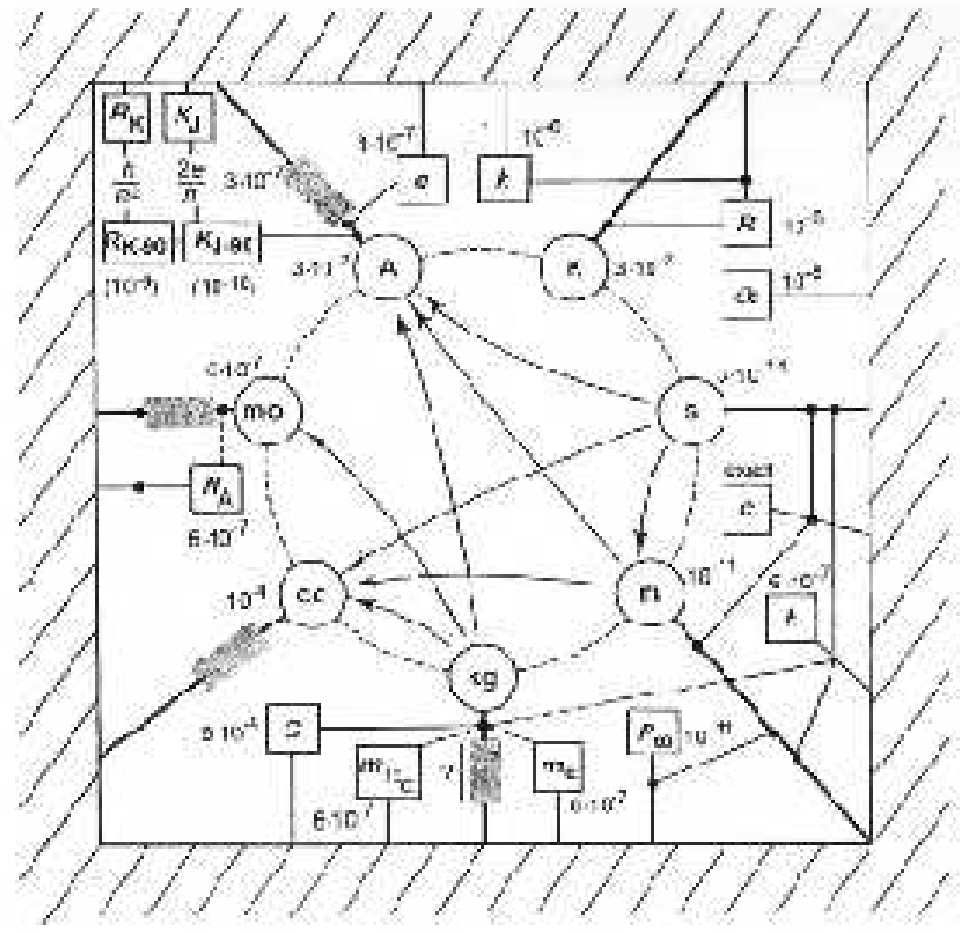}
\caption{The base units of the SI, with their present
uncertainties of realization, and some of their links to atomic
and fundamental constants with their present uncertainties in
terms of the SI. The absence of a useful quantitative estimate of
the long-term stability of the kilogram, indicated by ?, is
reflected in three of the other base units. The dashed lines to
the kilogram indicate possible routes to a new definition. (From
the article by T.J.~Quinn ``Base Units of the Syst\`{e}me
International d'Unit\'{e}s, their Accuracy, Dissemination and
International Traceability'', Metrologia {\bf 31} (1994/95)
515-527.) \label{fig9}}
\end{center}
\end{figure}

\label{04_}
\end{document}